\documentclass{aa}  
\usepackage{graphicx}
\usepackage{txfonts}
\usepackage{natbib}
\usepackage{rotating,hyperref}
\bibpunct{(}{)}{;}{a}{}{,} 
\usepackage{color,hyperref}
\newcommand{\etal}{\textit{et~al.}\/}
\newcommand{\ie}{\textit{i.e.}\/}
\newcommand{\eg}{\textit{e.g.}\/}
\newcommand{\cf}{\textit{\it cf.}\/}

\begin{document}
\title{The On The Fly Imaging Technique}

\author{J.~G.~Mangum,
        \inst{1}
        D.~T.~Emerson,
        \inst{2}
        and
        E.~W.~Greisen
        \inst{3}
        }
      
\offprints{J. Mangum}

\institute{National Radio Astronomy Observatory, 520 Edgemont Road,
           Charlottesville, VA  22903, USA \\
           \email{jmangum@nrao.edu}
           \and
           National Radio Astronomy Observatory, 949 North Cherry
           Avenue, Tucson, AZ  85721, USA \\
           \email{demerson@nrao.edu}
           \and
           National Radio Astronomy Observatory, P.O. Box O, 1003
           Lopezville Road, Socorro, NM  87801, USA \\
           \email{egreisen@nrao.edu}
           }

\date{Received 8 May 2007 / Accepted 31 Aug 2007}

 
  \abstract
   {}
   {The On-The-Fly (OTF) imaging technique enables single-dish radio
telescopes to construct images of small areas of the sky with greater
efficiency and accuracy.}
   {This paper describes the practical application of the OTF imaging
technique.  By way of example the implementation of the OTF imaging
technique at the NRAO 12 Meter Telescope is described.}
   {Specific requirements for data sampling, image formation, and
Doppler correction are discussed.}
   {}


   \maketitle
%

\section{Introduction}
\label{intro}

In an effort to become scientifically more efficient, astronomical
observatories have incorporated innovative observing techniques to
increase the throughput of their telescopes.  The On-The-Fly (OTF)
imaging technique is an innovation which has been adopted at a number
of radio observatories over the past forty years in a variety of
forms.  In the 1960's and 1970's, radio observatories which operate at
centimeter wavelengths implemented the ``drift scanning'' technique
(sometimes referred to as ``nodding'') to acquire continuum imaging
measurements (\cf\ \citet{Haslam1970}).  Drift scanned measurements
are obtained by slowly slewing the telescope in elevation with the
Earth's rotation providing the second dimension.  In the 1980's, a
number of radio telescopes added beam-switched continuum imaging
capabilities.  Beam-switched continuum observations are acquired by
slewing the telescope in a two-dimensional raster pattern over a given
region of sky while rapidly wobbling the telescope's subreflector to
obtain differential measurements of the total power.  The
\citet{Emerson1979} [EKH] algorithm 
allowed the deconvolution of these beam-switched measurements into
total power images. This powerful imaging capability added considerably
to our ability to understand the centimeter- and millimeter-wave
continuum structure of the interstellar medium.  In the 1990s a number
of radio observatories implemented both spectral line and continuum
OTF imaging capabilities (\cf\ \citet{Mangum1999} and
\citet{Mangum2000}).

\begin{figure}
\resizebox{\hsize}{!}{
  \includegraphics[scale=0.65]{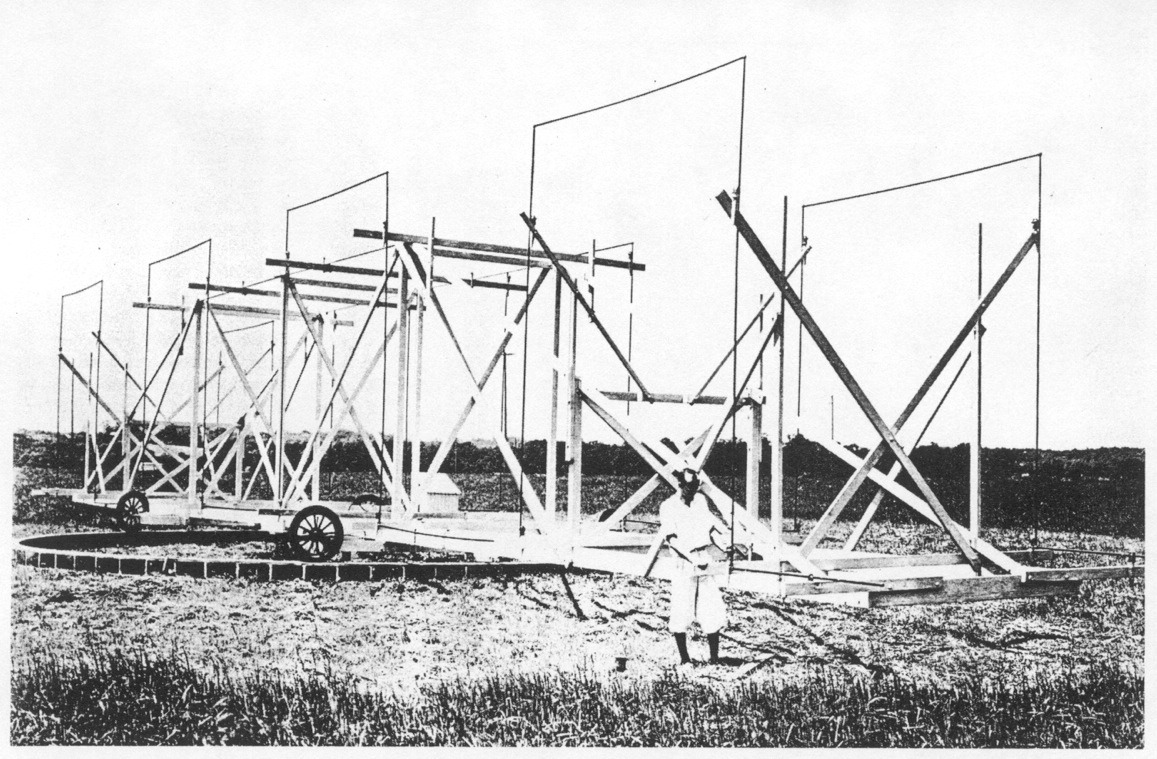}}
  \caption{Jansky and his telescope (\citet{Sullivan1978}).}
  \label{fig:jansky}
\end{figure}

As with many innovations, this ``new'' OTF imaging technique was just
an adaptation of an old observing technique.  The observations which
signaled the birth of radio astronomy, made by Karl Jansky in 1932,
were actually OTF observations (see \citet{Jansky1932} and
Figure~\ref{fig:jansky}).  Jansky's 20.5  
MHz synchrotron measurements of the galactic center represent the
first astronomical OTF observations.  In the following, we describe the
adaptation of Jansky's imaging technique to total power images
obtained using a single dish telescope.

\section{Advantages of OTF Observing}

OTF mapping is an observing technique in which the telescope is driven
smoothly and rapidly across a region of sky, or ``field'', while data
and antenna position information are recorded continuously.  This
technique is in contrast to traditional mapping of discrete positions
on the sky, which is sometimes called ``step-and-integrate'' or
``point-and-shoot'' mapping.  The advantages to OTF mapping are:

\begin{enumerate}

\item Telescope overhead is reduced significantly, since a specific
  position on the sky does not have to be acquired within a given
  tolerance (\ie\ minimizing ``dead-time'').

\item The entire field is covered rapidly, minimizing changes in the
  properties of the atmosphere and the system, including antenna
  pointing and calibration.  Systematic changes may occur from map to
  map, but such effects average down rapidly and may be correctable by
  cross correlation techniques.  In general, global changes from map
  to map are more benign and easier to correct than drifts across a
  single map field.

\item Higher observing efficiency:  The radiometer equation for
switched measurements is given by the following

\begin{equation}
\sigma = \frac{T_{sys}}{\eta_{spec}\sqrt{\Delta\nu
t_{on}}}\left[1 + \frac{t_{on}}{t_{off}}\right]^\frac{1}{2}
\label{eq:radiometer}
\end{equation}

where $\sigma$ is the RMS noise in a total power measurement,
T$_{sys}$ is the system noise temperature, $\Delta\nu$ is the spectral
resolution of the measurement, and $\eta_{spec}$ is the spectrometer
efficiency.  The optimum duration of an OFF measurement for any switched
measurement is given by (\citet{Ball1976})

\begin{equation}
t^{optimal}_{off} = \sqrt{N} t_{on}
\end{equation}

where N is the number of ON measurements made per OFF measurement,
such that $t_{scan} = t_{off} + Nt_{on}$.
Therefore, for optimal observing the RMS noise in a measurement
is given by

\begin{equation}
\sigma = \frac{T_{sys}}{\eta_{spec}\sqrt{\Delta\nu
t_{scan}}}\left[1 + \frac{1}{\sqrt{N}}\right]
\end{equation}

where $t_{scan}$, as defined above, is the total ON plus OFF
measurement time.  For N$\geq$100, $\sigma$ is approximately a factor
of two 
smaller than it is for an equivalent position switched measurement,
which leads to a factor of four improvement in observing time
efficiency.

\end{enumerate}

\section{Sampling}
\label{sampling}

\subsection{Detection and Sampling: The Spectrum of the Detected Noise
  and Signal} 
\label{spectrum}

\begin{figure}
\resizebox{\hsize}{!}{
  \includegraphics[scale=0.85]{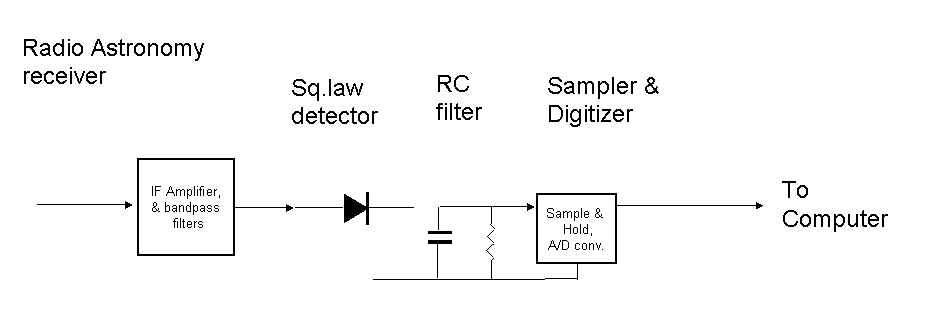}}
  \caption{Typical continuum detection scheme for a single dish
    telescope.}
  \label{fig:contdetect}
\end{figure}

When conducting single dish imaging observations, it is important to
keep in mind the following facts about sampling and aliasing in radio
astronomical mapping data.  If you want to represent the full
resolution of the telescope, you have to sample the data often enough
to represent all the spatial frequencies detected by the antenna.
Figure~\ref{fig:contdetect} illustrates a typical continuum detection
scheme of a single dish telescope; this is the simplest case, but the
same principles apply to detection in a spectrometer, or even direct
digitization at the IF and subsequent processing by software or
firmware.  In the example shown, the radio astronomical signal is
amplified and bandpass filtered, before passing into a square law
detector, possibly some simple filtering, and then into an
Analog-to-Digital (A/D) converter.  The A/D converter, or later
processing, usually incorporates some signal integration before
delivering the sampled data.

\begin{figure}
\resizebox{\hsize}{!}{
  \includegraphics[scale=0.85]{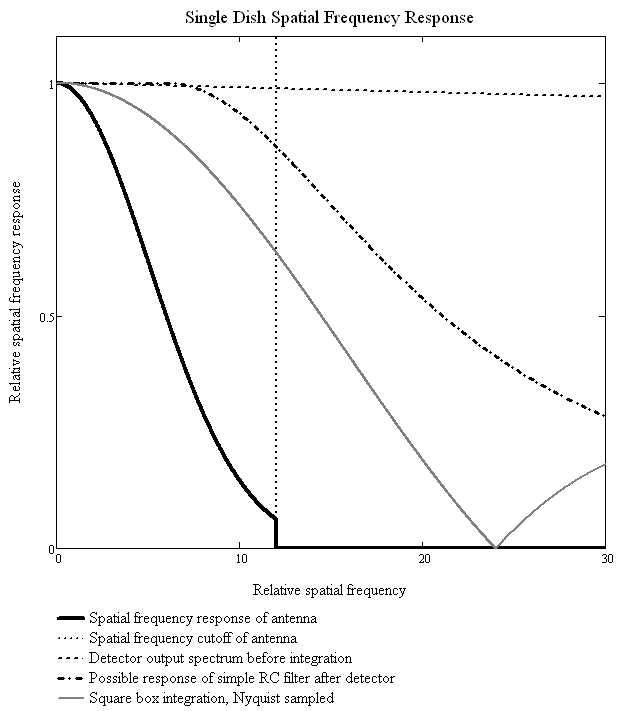}}
  \caption{Illustration of the low-frequency spectral response of the
    signal and noise after square law detection.}
  \label{fig:lowfrequencyspec}
\end{figure}

Figure~\ref{fig:lowfrequencyspec} illustrates the low frequency
spectrum of the signal and noise immediately after the square law
detector and in the subsequent processing steps.  With the antenna
scanning at a constant speed over a point source, its response
produces, after the detector, a time-varying signal whose spectrum
corresponds to the spatial frequency response of the antenna.  The
horizontal axis of Figure~\ref{fig:lowfrequencyspec} could be spatial
frequency, measured in units proportional to $\frac{d}{\lambda}$ (d 
is a baseline distance), or the frequency after the detector measured
in Hz.  Any antenna has a spatial frequency cut-off; with an antenna
of diameter D and scanning speed of $\Omega$ radians/second, after the
detector the signal voltage will have a cut-off frequency of
$\left(\frac{D}{\lambda}\right)\Omega$, now measured in Hz.  This is
illustrated by the bold curve in Figure~\ref{fig:lowfrequencyspec}.
The signal cut-off frequency is shown by the vertical dashed line.

Figure~\ref{fig:lowfrequencyspec} also shows the spectrum of noise
(the dashed line) immediately after the detector; this is strictly the
autocorrelation function of the IF passband, which means in practice
that within the narrow range of post-detection frequencies of
interest, it is white noise, with a flat spectrum.  It is normal to
include a simple low pass filter immediately after the detector, shown
symbolically by the RC time constant in Figure~\ref{fig:contdetect}.
This might have a gentle frequency cut-off such as illustrated by the
dash-dot line of Figure~\ref{fig:lowfrequencyspec}.

The digitizer usually includes some time integration as part of its
operation; the output samples might be sampled S times per second, but
each sample is an integration of $\frac{1}{S}$ seconds of data.  This
integration is equivalent to convolving the time sequence of data with
a square-box function -- equivalent also to convolving the sky image
along the scanning direction by the corresponding square-box function.
The integration of S seconds by this square-box translates into an
average of $S*\Omega$ radians of angle along the sky scanning track.
The time-domain spectral response, or equivalently the spatial
frequency response, of this square-box integration is illustrated in
Figure~\ref{fig:lowfrequencyspec} for the case where the signal is
exactly Nyquist sampled (two independent samples per
$\frac{\lambda}{D}$ interval on the sky). 

The \textit{ideal} signal processing and detection scheme might use a perfect
square-edged anti-aliasing filter in place of the simple RC-filter
illustrated in Figure~\ref{fig:contdetect}, followed by a sampler
operating at least at twice the spatial frequency cut-off, with an
infinitely narrow window in the sampler, thereby avoiding the loss of
high frequencies caused by the normal square box integration.
However, to avoid the complexities (and impossibility) of building a
perfect anti-aliasing filter, very often in practice the square-box
integration is retained, but with a substantial increase in the sample
rate above the Nyquist value.

\subsection{The Consequences of Undersampling}
\label{undersampling}

The spectrum of noise and of signal at the output of the detector of a
total power system depends on the details of the electronic design,
but a representative system is illustrated in
Figures~\ref{fig:contdetect} and \ref{fig:lowfrequencyspec}.

With OTF observing, the astronomical sky is convolved with the
telescope beamshape.  The telescope acts also as a low pass filter,
removing all astronomical components beyond the intrinsic angular
resolution of the antenna.  The telescope beam and telescope scanning
speed together give a low pass filter, with the conversion from
spatial frequency to temporal frequencies being given by

\begin{equation}
R(f)=r(sf)*\Omega,
\label{eq:spatialfreq}
\end{equation}

\noindent{with} R(f) being the temporal frequency response in
Hz and r(sf) the spatial frequency response measured in units of
$\frac{d}{\lambda}$, with $\Omega$ the telescope scanning speed in
radians/second.  The respective responses are illustrated in
Figure~\ref{fig:lowfrequencyspec}.

\subsubsection{Aliasing}
\label{aliasing}

The minimum sampling interval on the sky, for an antenna with maximum
dimension D, is at intervals of $\frac{\lambda}{2D}$.  Assume that
one undersamples on the sky, rather than later in the data 
processing.  Suppose you have a 10m dish, but you only sample at
$\lambda/(2\times 8~m)$ rather than the $\lambda/(2\times 10~m)$
that you should.  This means that the spatial frequencies 
present from the dish baselines of 8~m to 10~m get reflected back 
into the spatial frequencies of 8~m down to 6~m.  Not only have spatial
frequencies from the 8~m to 10m baselines been lost, but valid spatial 
frequencies from baselines of 6~m to 8~m have been corrupted.  You can't
tell if structure in your map with a spatial wavelength of
$\lambda/7~m$ is genuine, or was really structure at
$\lambda/9~m$ which has been aliased on top of any genuine
$\lambda/7~m$ spatial wavelength signal.  In this sense,
undersampling the sky is really twice as bad as you might have thought.

Figures~\ref{fig:spatial} and \ref{fig:aliased} show the spatial
frequency response and aliased noise power for Nyquist sampling while
using a square box sampling function, of width equal to the sampling
interval. The power P$_s$ in the random noise fluctuations in
frequencies up to the properly sampled spatial frequency, indicated by
the vertical dashed line at 0.5 on the abscissa of
Figure~\ref{fig:spatial}, is proportional to:

\begin{figure}
\resizebox{\hsize}{!}{
\centering
  \includegraphics[scale=0.5,angle=-90]{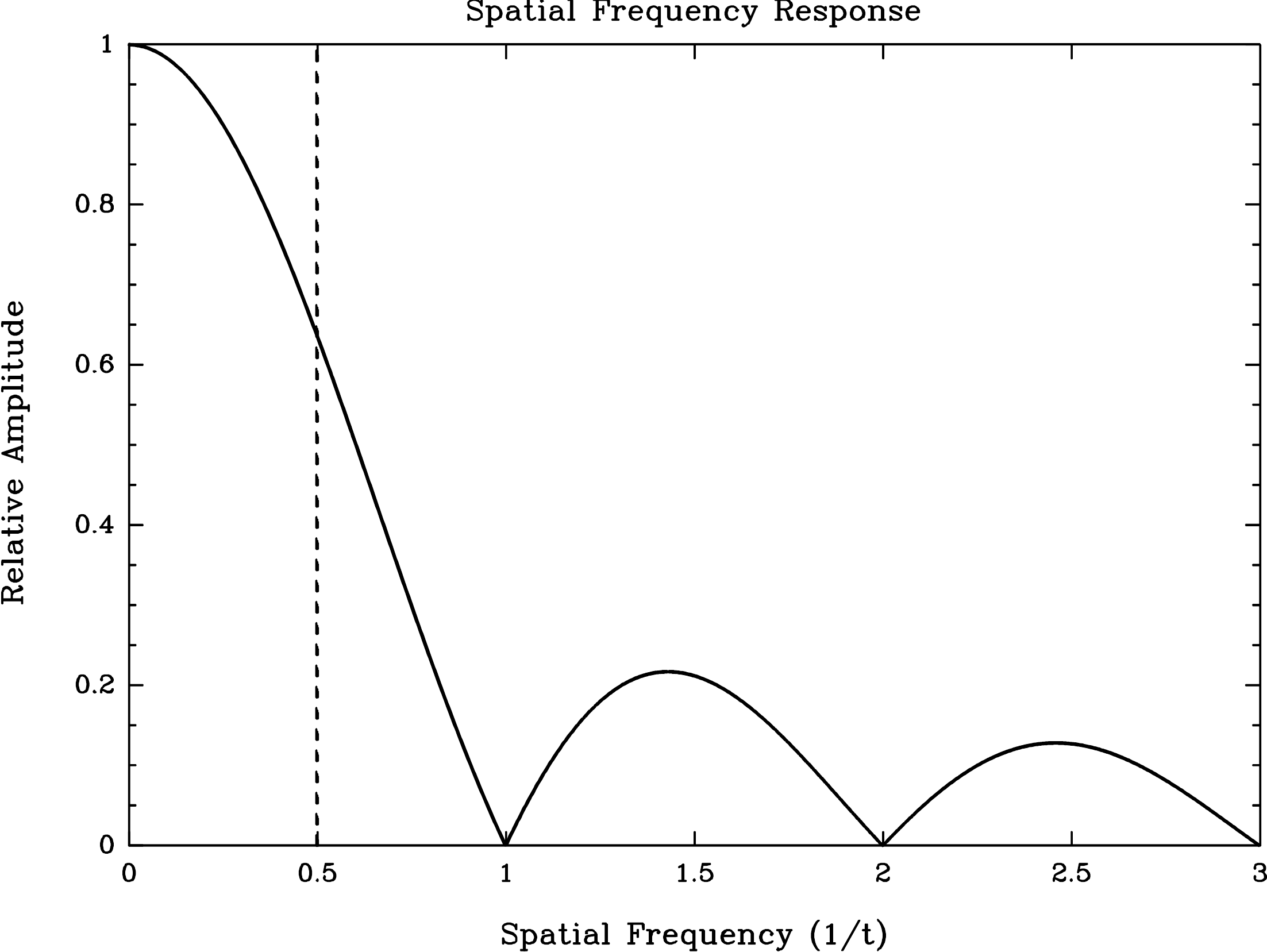}}
  \caption{Spatial frequency response for square box Nyquist sampling.}
  \label{fig:spatial}
\resizebox{\hsize}{!}{
  \includegraphics[scale=0.5,angle=-90]{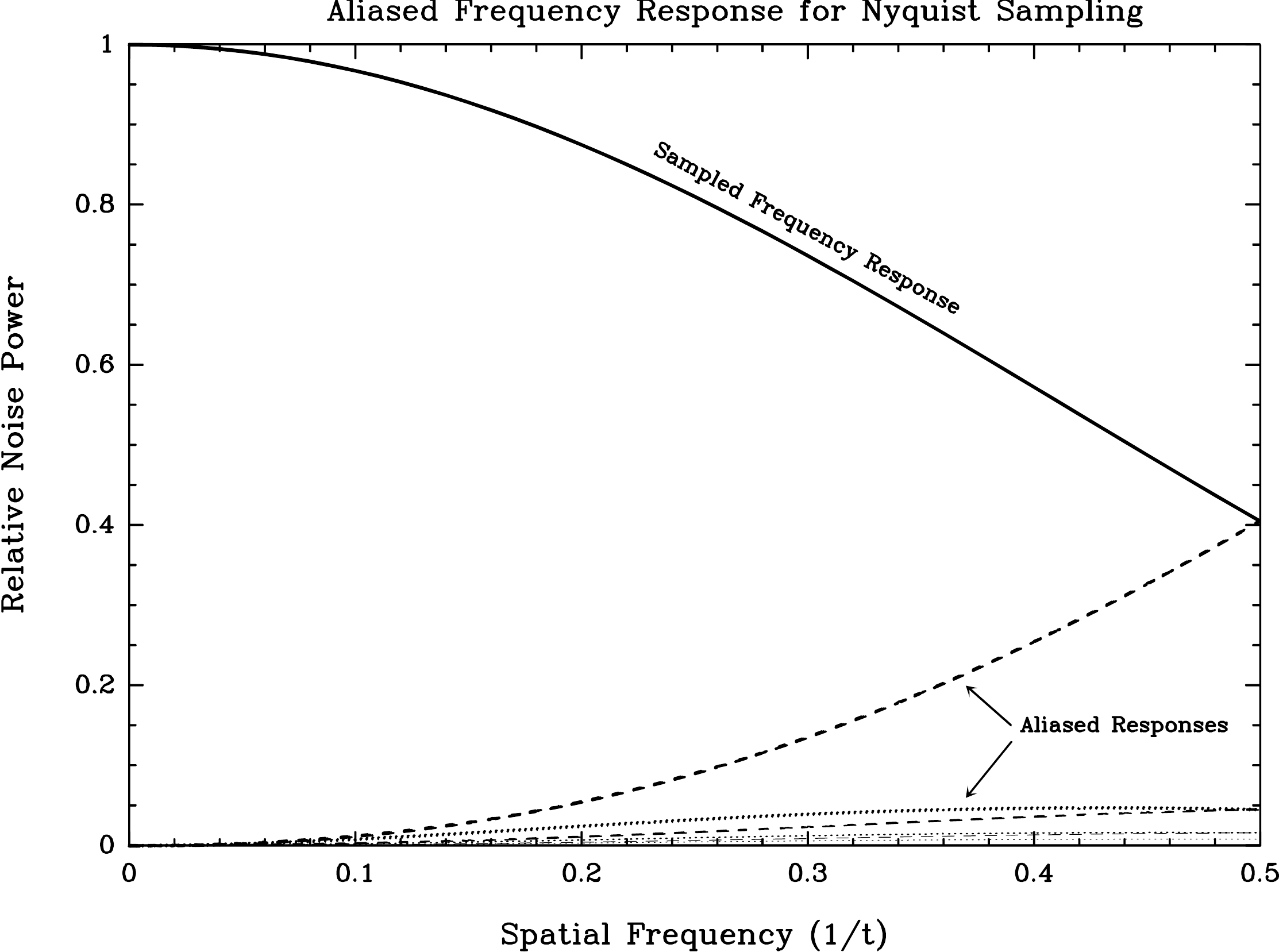}}
  \caption{Aliased noise power for square box Nyquist sampling in the
    absence of a separate anti-aliasing filter.}
  \label{fig:aliased}
\end{figure}

\begin{equation}
P_s = \int_0^{0.5} \left(\frac{\sin(\pi x)}{\pi x}\right)^2 dx
\label{eq:ps}
\end{equation}

The power A$_n$ in the random fluctuations in frequencies beyond that
frequency, in absence of any anti-aliasing or other filters, is
proportional to:

\begin{equation}
A_n = \int_{0.5}^{\infty} \left(\frac{\sin(\pi x)}{\pi x}\right)^2 dx
\label{eq:an}
\end{equation}

\noindent{This} noise power A$_n$, which cannot contain any
astronomical information, is then aliased back into the signal
passband below 0.5. The random (RMS) noise voltage is then increased
by the ratio:

\begin{equation}
\sqrt{\frac{A_n + P_s}{P_s}} = 1.137
\label{eq:sqrtanps}
\end{equation}

\noindent{equivalent} to a loss in signal-to-noise of 12\%.
Therefore, the noise at frequencies to the right of 0.5 on the abscissa of
Figure~\ref{fig:aliased}, aliased back into the region of interest to
the left of 0.5, leads to an overall degradation in signal-to-noise of
the image of 12\%.   

\begin{figure}
\resizebox{\hsize}{!}{
  \includegraphics[scale=0.5,angle=-90]{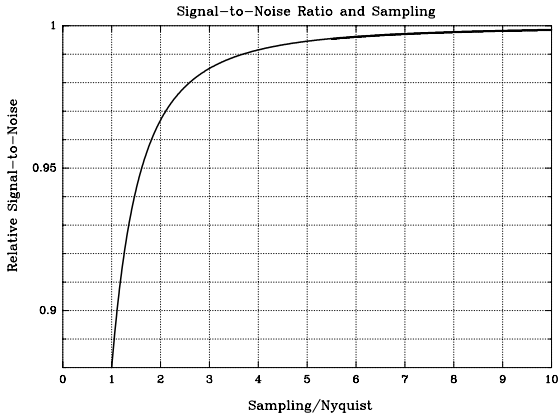}}
  \caption{Signal-to-noise degradation due to noise aliasing of an
  inadequately sampled image using square box sampling.
}
  \label{fig:snr}
\end{figure}

The noise power degradation due to sampling is a function of the level
of oversampling in the image plane.  Figure~\ref{fig:snr} shows the
amount of signal-to-noise degradation one can expect due to noise
aliasing of an inadequately sampled signal using square box sampling.

How important undersampling is depends on exactly what the
illumination taper is, how 
important it is that you retain the maximum possible resolution of the 
telescope, how good a dynamic range you want in the observations, and
at some level how much fine scale structure there is in the source itself.
If you only sample at 0.833 Nyquist (\eg\ $FWHM/2$ rather than
$FWHM/2.4$ for a $-13$ dB illumination taper), what matters is the
energy in the data at spatial wavelengths shorter than
$\lambda/(2\times 0.8\times D)$.  So in
a sense you need to ask what the illumination taper is at a radius of
$0.4\times D$ on the dish surface.  The spatial frequency
response of a single dish is the autocorrelation function of the 
voltage illumination pattern.  So, you need to calculate how much
area there is under the 2-D autocorrelation function beyond spatial
frequencies of $0.8\times D$, compared to the area within $0.8\times
D$.  This ratio is some measure of the dynamic range.  A better
definition of dynamic range might take into account the spatial
frequency structure of the source.  If the source has no structure on
scales smaller than $\lambda/(2\times 0.8\times D)$, then you
don't need to sample at the full $\lambda/(2\times D)$ anyway.

There are circumstances where it is perfectly rigorous to undersample the
data in the absence of a separate anti-aliasing filter.  For example,
if you have a 10 m dish, and you are taking data 
to compare with other observations that were made using a 1 m dish at
the same wavelength (or the 
equivalent number of wavelengths at some other frequency) then you only
need to sample the data at $\lambda/(2\times 5.5~m)$ or
$\lambda/11~m$.  This is so because, when
sampling a 10~m dish as if it were a 5.5~m dish, the spatial frequency
components from baselines of 5.5~m out to 10~m will be reflected back
into the data as if corresponding to baselines of 5.5~m down to 1~m.
So, the spatial frequency terms of the 1~m baseline and below will not
have been 
corrupted.  The data analysis of this undersampled data would apply a
spatial frequency cutoff at 1~m, and there will have been no corruption
in this smoothed data caused by the undersampling.  Putting it in more general 
terms, if you are going to be smoothing observations from a dish of
diameter $D$ to simulate observations made with a smaller dish of
diameter $d$, then the sampling interval only needs to be
$\lambda/(d+D)$ rather than $\lambda/(2\times D)$.

\subsubsection{Beam Broadening}
\label{beambroadening}

Since the telescope beam is effectively convolved with a square-box
function equal to the distance the telescope moves in one sample
period, there is a degradation of spatial resolution due to the data
acquisition process.  If you set an arbitrary criterion that the
telescope beam shouldn't be broadened by more than 1\%, then you need
to sample at least 4.5 points per FWHM beam width.  Note that since
Nyquist sampling is about 2.4 points per FWHM, assuming an
illumination taper of about $-13$ dB; for a uniformly illuminated
circular aperture the factor would be about 2.06, while for a $-20$ dB
taper it becomes 2.60.  This means that one needs to
sample at about twice Nyquist to avoid degradation of resolution worse
than 1\%.

\subsubsection{Sampling Summary}
\label{sampsummary}

\begin{enumerate}

\item To avoid both beam degradation and loss of signal/noise ratio,
  one must sample at least twice Nyquist, or about 5 points per
  FWHM.

\item With few exceptions, undersampling at data acquisition is not
  recommended.

\item There are other aspects that make it desirable to sample
  \textbf{more} often than the Nyquist rate, as is often recommended
  for OTF observing at many millimeter-wave telescopes.  These are
  practical points like how well gridding or interpolation works with
  a finite sized gridding or interpolation function.  For example, a
  little oversampling may enable you to reduce the convolution
  (interpolation) function by a factor of a few, saving a huge amount
  of computational overhead at the expense of a few per cent more
  data.

\end{enumerate}

\section{Gridding (Convolution)}
\label{gridding}

\begin{figure}
\resizebox{\hsize}{!}{
  \includegraphics[scale=0.50,angle=-90]{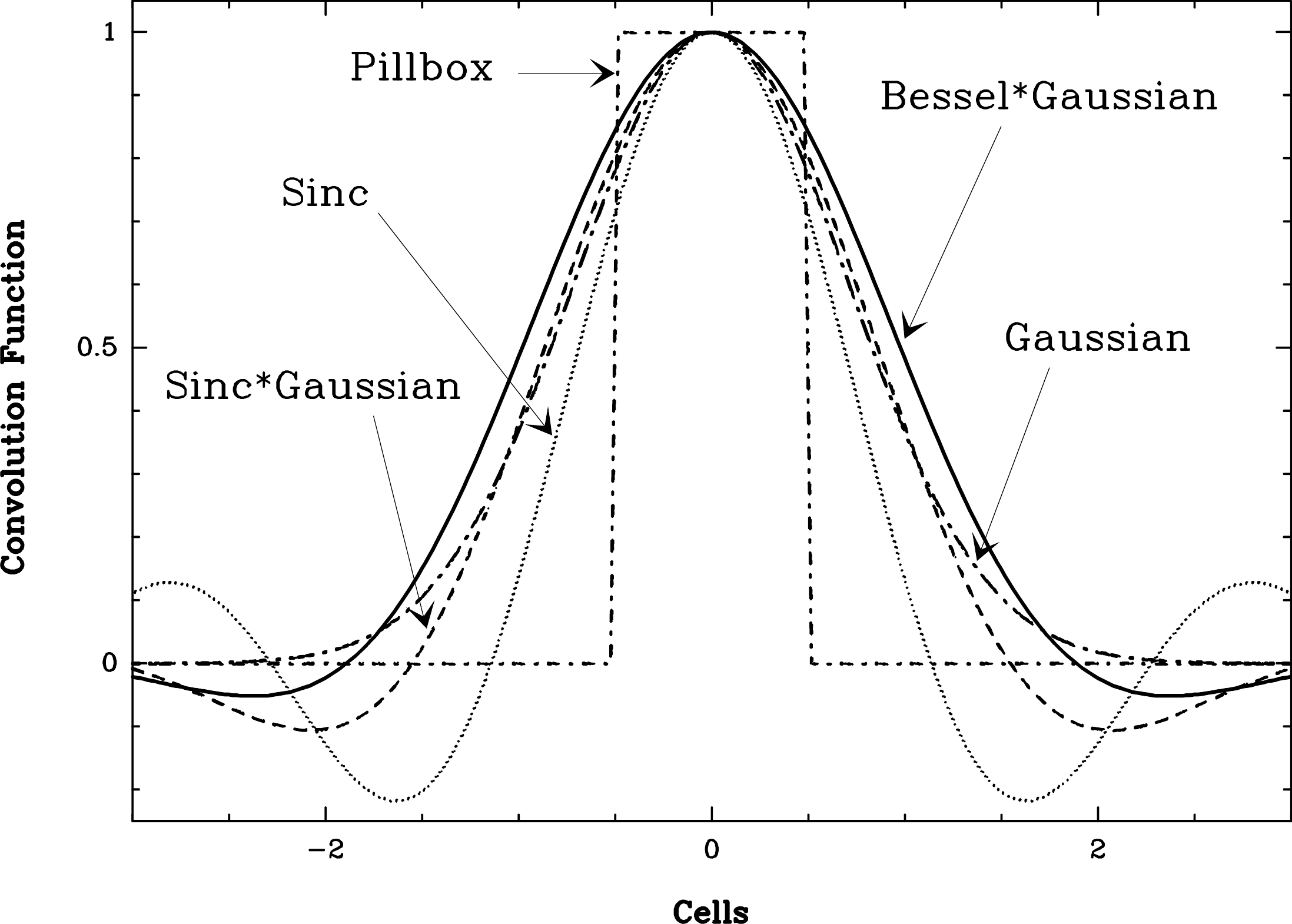}}
  \caption{Representative convolution functions.}
  \label{fig:confunc}
\end{figure}

\begin{figure}
\resizebox{\hsize}{!}{
  \includegraphics[scale=0.50,angle=-90]{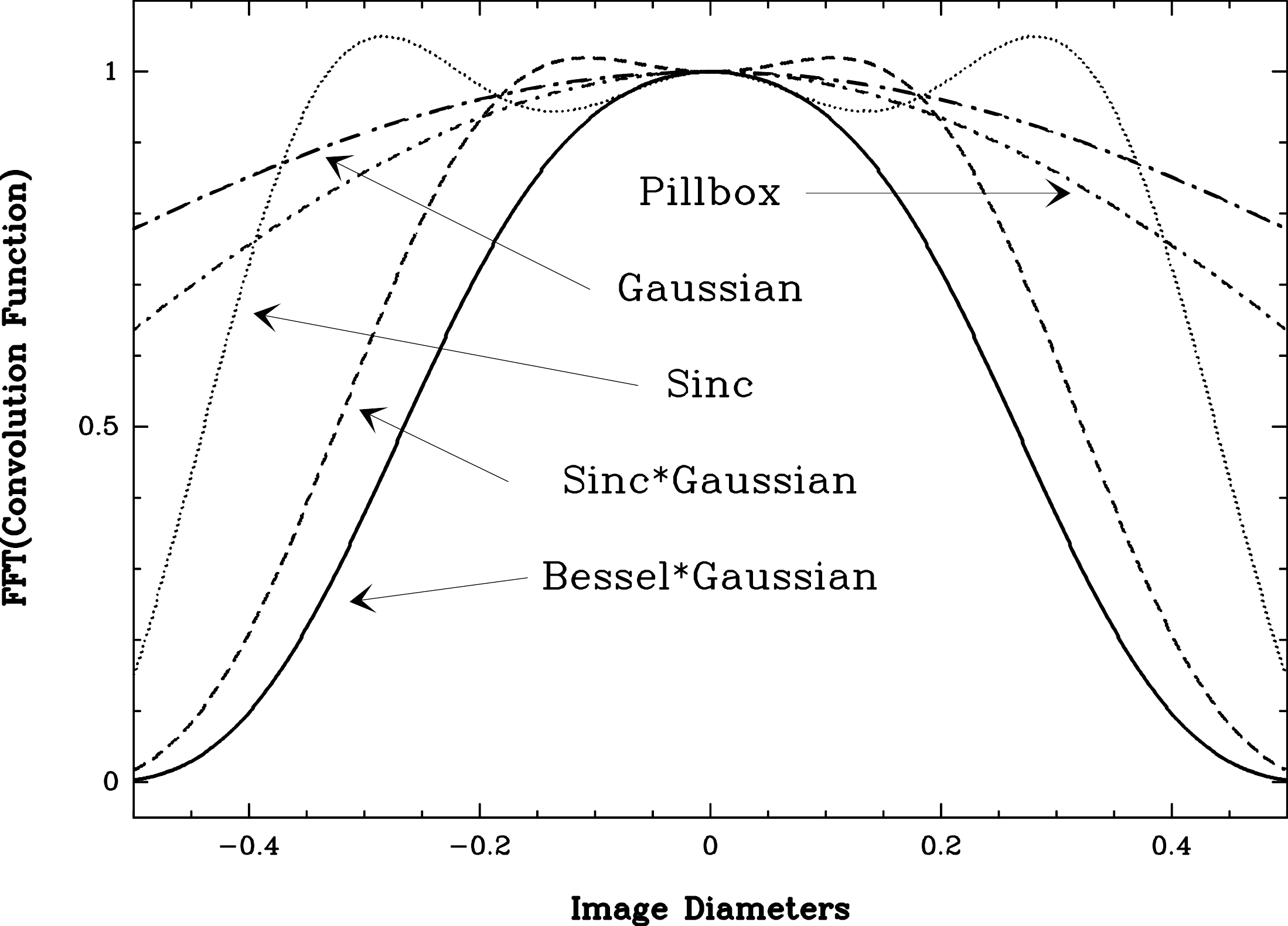}}
  \caption{Representative Fourier transforms of convolution functions.}
  \label{fig:fftfunc}
\end{figure}

The final image is constructed by smoothing the data at whatever
coordinates they were observed and then re-sampling them on a regular
image grid.  The smoothing is, in fact, an interpolation, which can
include weighting and averaging, rather than a strict convolution.
However, when the data are sampled in a dense and nearly uniform
fashion, the interpolation approximates a convolution.
Figure~\ref{fig:confunc} shows some representative convolving
functions which may be used in the data gridding.  The effect of the
convolution is best seen by examining its impact, a multiplication, in
the Fourier, spatial frequency, space.  The Fourier transforms of the
convolving functions are illustrated in Figure~\ref{fig:fftfunc}.

The spatial frequency response pattern of the single dish is
effectively multiplied by the Fourier transform of the convolving
function.  If the function were to have no effect on the data, then
its Fourier transform should have value 1.0 out to the maximum spatial
frequency of the telescope and 0.0 outside that radius.
This function is, in one dimension, a sinc function, $\sin (\pi x /
NS) / (\pi x / NS)$ and, in two dimensions, a Bessel 
function $ J_1(\pi x / NS) / (\pi x / NS)$.  Unfortunately, both of
these functions 
die off very slowly, making the convolution operation excessively
expensive.  A compromise function is obtained by tapering the
convolution function by a fairly narrow Gaussian, allowing the
function to be truncated outside a restricted region.  In the Fourier
space, this has the effect of reducing the response to the largest
spatial frequencies available to the telescope.  These spatial
frequencies are probably already tapered by the response pattern of
the feed horn, which is usually designed to avoid sensitivity to
ground spillover.

Since we have a circular telescope and are gridding the image in two
dimensions simultaneously, we use the circularly symmetric version of
the convolving function, $J_1(r/a) / (r/a) exp (-(r/b)^c)$.
Optimizations of this function give the default values for a, b, and c
of a = 1.55 (single-dish-beam-size / 3), b = 2.52
(single-dish-beam-size / 3), c = 2 (see Table~\ref{tab:convfunc}), and
a support radius equal to the single-dish-beam-size (FWHM)
(\citet{Schwab1980}).

It is important to choose the sampling of the output image, and the
width and support of the convolving function to avoid aliasing.  This
topic was discussed at length in \S\ref{aliasing}.  A suitable cell
size would be less than or equal to one-third of the
single-dish-beam-size.

Other convolving functions have various undesirable effects.  For
example, a $sinc^2$ has a triangular Fourier transform.  All non-zero
spatial frequencies are weighted down linearly with frequency,
degrading the telescope resolution.  All other positive-only functions
have a similar effect.  One can construct convolving functions such 
as $\sin (x/a) \sin (y/a) / (x/a) / (y/a) exp (-((x/b)^2 + (y/b)^2))$
which will actually improve the spatial resolution of the image over
that of the single-dish beam.  They do this at the cost of
down-weighting low spatial frequencies and thereby reducing the
signal-to-noise of the observation (\citet{Greisen1998}).


\section{Scanning Geometry}
\label{scanning}

The scanning geometry used to acquire an OTF map is usually dependent
upon the type of source you are imaging and the ability to command the
antenna to execute complex patterns.  Scanning geometries that have
been implemented at existing radio telescopes include:

\begin{description}
\item[Raster:] Usually the first scanning geometry implemented at a
  telescope as it is the simplest to implement within a telescope
  monitor and control system.  Figure~\ref{fig:raster} is a pictorial
  display of this scanning geometry.

\begin{figure}
\resizebox{\hsize}{!}{
  \includegraphics[scale=0.3]{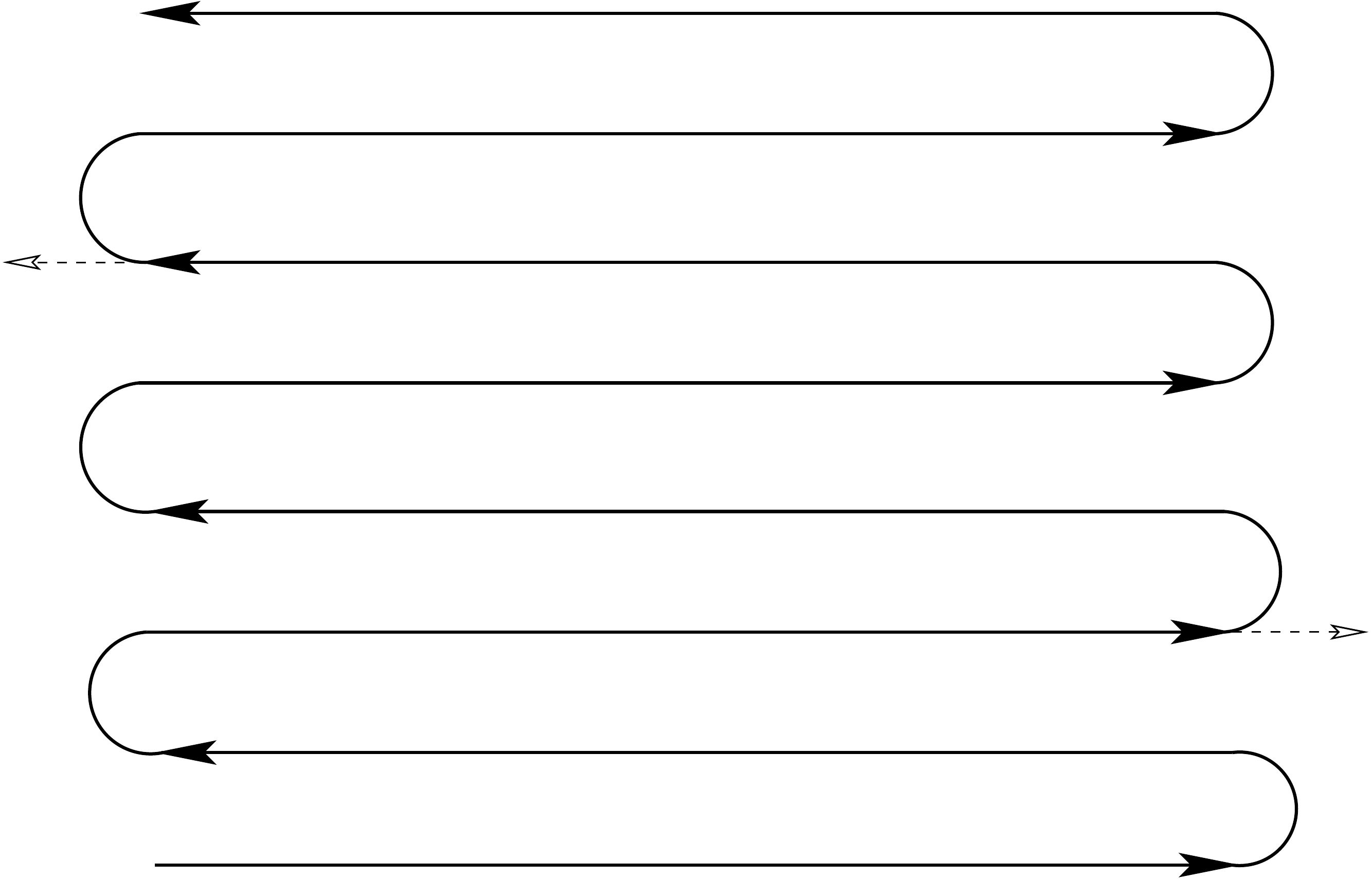}}
  \caption{An example of a raster scanning pattern.  In this example,
  three scanning rows are acquired per reference (``off'')
  measurement.} 
  \label{fig:raster}
\end{figure}

\item[Spiral:] An efficient scanning geometry when imaging a circular
  or elliptical source (see Figure~\ref{fig:spiral}).

\begin{figure}
  \centering
  \includegraphics[scale=0.4]{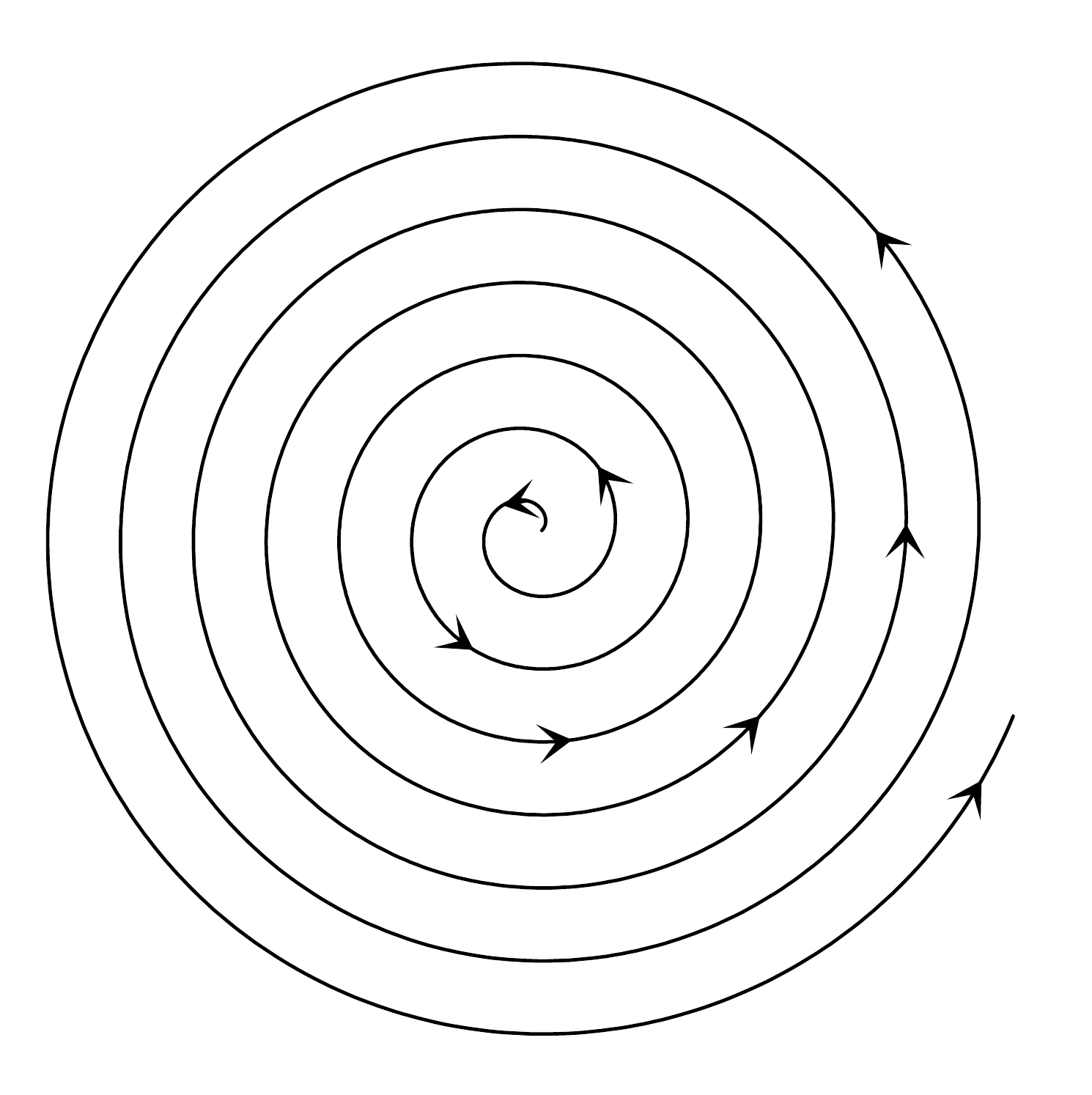}
  \caption{An example of the spiral scanning pattern.}
  \label{fig:spiral}
\end{figure}

\item[Hypocycloid:] For imaging a non-circular source which is large
  in relation to the resolution element of your observations, the
  hypocycloid is a very motion-efficient pattern (\ie\ no antenna
  motion efficiency losses due to turn-around; see
  Figure~\ref{fig:hypocycloid}).  

\begin{figure}
  \centering
  \includegraphics[scale=0.6]{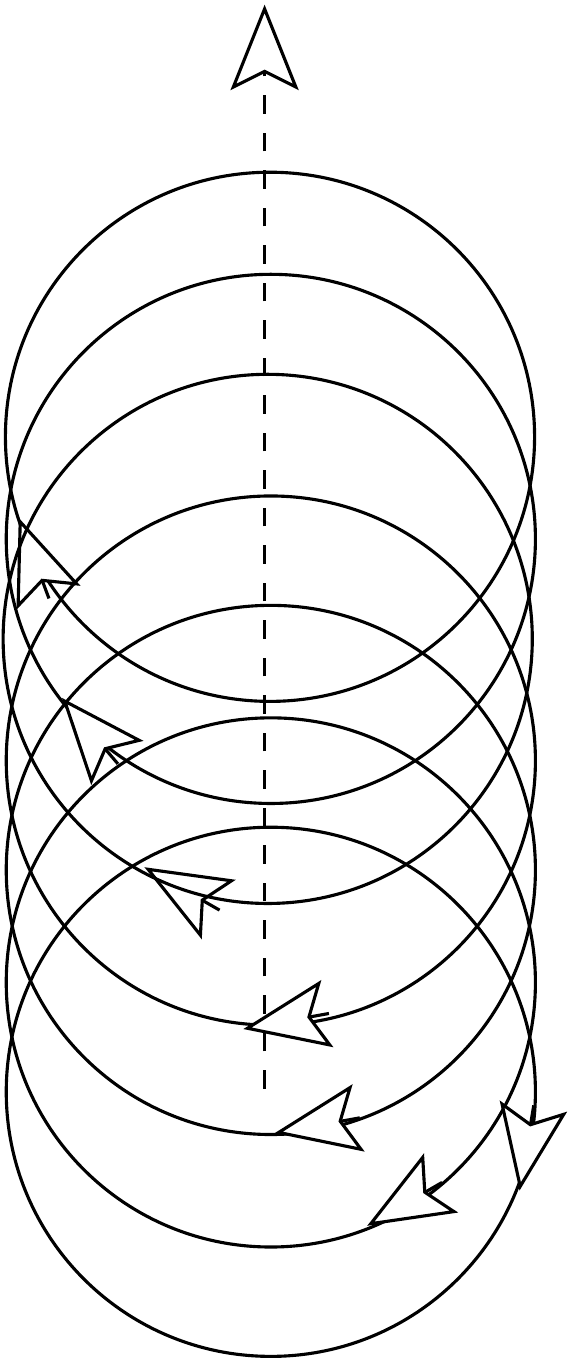}
  \caption{An example of the hypocycloid scanning pattern.}
  \label{fig:hypocycloid}
\end{figure}

\end{description}

\section{OTF Imaging in Practice}

It is often important to estimate the amount of integration time and
the RMS noise level an OTF image will have based on the observing
parameters.  In the following, we derive these quantities for a raster
scanned OTF map.  We also discuss an often overlooked aspect of
wide-field spectroscopy involving the doppler correction.

\subsection{OTF Map Parameters}

In the following we describe some of the OTF map parameters that one
must generally calculate in order to setup an OTF observation.

\subsubsection{Row Sampling Rate}
\label{rowsamp}

It is very important that you set up your map to be properly sampled
in all coordinates.  If you under sample, you will
miss information in the image field, you will be unable to combine
your image accurately with that from interferometers or other single-dish
telescopes, and you may introduce artifacts from
the analysis algorithms (see \S\ref{aliasing}).  Keep in mind that you
can always smooth the map after it is taken to degrade the resolution
and improve signal-to-noise. 

The scanning rows must be spaced at no more than the Nyquist spacing,
which is also known as ``critical sampling'' and is given by 
  
\begin{eqnarray}
\theta_N &\equiv& \frac{\lambda_{obs}}{2 D}~\mathrm{radians} \nonumber \\
         &\simeq& \frac{30918}{D(m) \nu_{obs}(GHz)}~\mathrm{arcsec}
\label{eq:thetanyq}
\end{eqnarray}

\noindent{In} practice, though, a small amount of oversampling is
recommended.  If the rows are critically spaced, small 
scanning errors can result in the map being under sampled.  In
addition, the tail of the gridding function used by the analysis
procedure extends slightly beyond the information cutoff of the
telescope, which will result in some noise being aliased into each
grid cell (see \S\ref{aliasing}).  A reasonable approach is to sample
both sky coordinates by:

\begin{eqnarray}
\theta_{row} &=& \left(\frac{\lambda_{obs}}{2D}\right)\times{0.9} -
             \delta \nonumber \\
             &\simeq& \left(\frac{27826}{D(m) \nu_{obs}(GHz)}\right)
             - \delta~\mathrm{arcseconds}
\label{eq:thetarow}
\end{eqnarray}

\begin{figure}
\resizebox{\hsize}{!}{
  \includegraphics[scale=0.35,angle=-90]{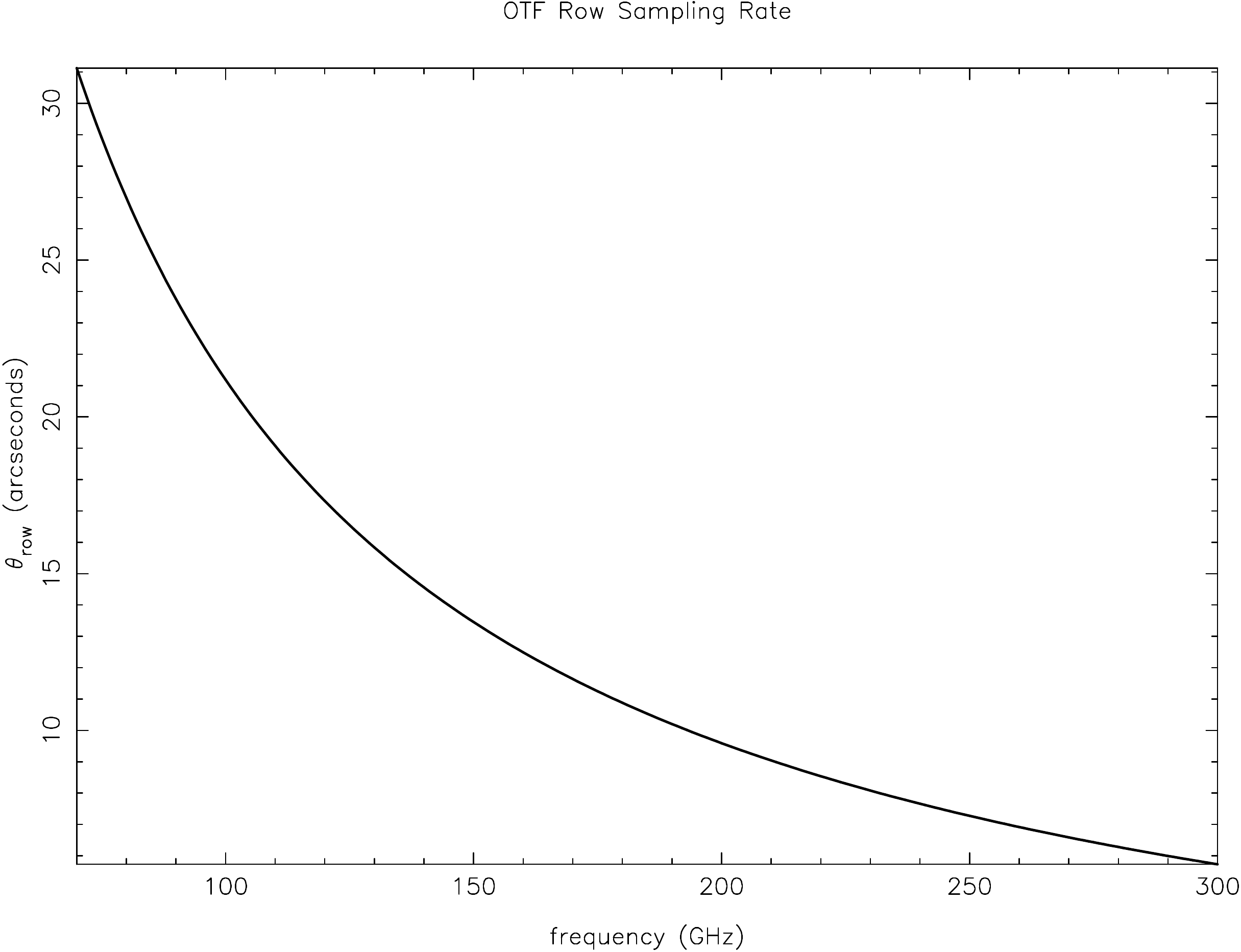}}
  \caption{$\theta_{row}$ as a function of observing frequency for a
  12m aperture.}
  \label{fig:samprat}
\end{figure}

\noindent{where} the factor of 0.9 is an oversampling factor and the
factor of $\delta$ is a guard band to accommodate any
scanning errors.  For the NRAO 12M Telescope $\delta = 2$ was used.
Figure~\ref{fig:samprat} is a plot of this relationship assuming a 12m
aperture.

\subsubsection{Scanning Rate}
\label{scanrat}

Map sampling along the scanning direction is dependent upon the
integration (or ``dump'') time of the data acquisition system (\ie\
spectrometer) being used.  Since the spectrometers must integrate for
a finite interval before being dumped, the data are ``square-box''
binned.  To avoid noise aliasing and beam-smearing problems due to
this binning (see \S\ref{aliasing}), one needs to oversample in the 
scanning direction.  The scan rate may be written as 

\begin{equation}
  R(arcsec/sec) = \frac{\theta_N}{n_{os} t_{dump}}
\label{eq:scanrat}
\end{equation}

\noindent{where} $\theta_N$ is the Nyquist spacing (Equation
\ref{eq:thetanyq}), $n_{os}$ is an oversampling factor, and $t_{dump}$
is the spectrometer data dumping interval.  At the NRAO 12 Meter
Telescope, $t_{dump}$ = 0.1 seconds for spectral line OTF, 0.25
seconds for continuum OTF.  The general recommendation to observers
when using the 12m system was to use a minimum value for $n_{os}$ of
2, which will result in $<$ 3\% increased noise as a result of
aliasing and minimal beam broadening (see \S\ref{beambroadening}). 
Note that it is quite 
acceptable to oversample by much larger factors than the minimum value
suggested above, particularly in the scanning direction.  There are
some tradeoffs to be considered, however:

\begin{enumerate}
\item  Large factors of oversampling will produce better
  signal-to-noise on a single coverage of the field and may result in
  simpler data processing, \ie\ fewer maps to average together.

\item On the other hand, the longer a single coverage takes, the more
  susceptible your image will be to drifts in the system, such as
  pointing and atmosphere.  This eliminates some of the advantages of
  fast mapping mentioned above.
\end{enumerate}

A good compromise is to use a row spacing computed from
Equation~\ref{eq:thetarow} and a scan rate calculated from
Equation~\ref{eq:scanrat}, with $n_{os} > 2$.  Other circumstances may
come into play, of course.  For example, if the field size is very
small, you may choose to slow the scan rate (increase $n_{os}$)
substantially.

\subsubsection{Map RMS}

When planning an OTF observing run you must calculate
how much integration time you need for each sampling cell to
reach the desired signal-to-noise level.  First determine
the integration time required using the standard Radiometer
Equation (Equation \ref{eq:radiometer})

\begin{equation}
  \sigma_{cell} = \frac{T_{sys}}{\eta_{spec}\sqrt{\Delta\nu
  t_{cell}}}\left[1 + \frac{t_{cell}}{t_{off}}\right]^\frac{1}{2}
\label{eq:radiometercell}
\end{equation}
  
\noindent{where} $T_{sys}$ is the system temperature, $\eta_{spec}$ is
the spectrometer efficiency (typically $\simeq 1$), $\Delta\nu$ is the spectral resolution,
$t_{cell}$ is the time spent integrating on a cell, and $t_{off}$ is
the off-source integration time.  For spectral line OTF, the time
spent integrating on a given cell is typically one second or less,
while the off-source integration time is often several seconds (10
seconds was the default value for the NRAO 12m Telescope).  Therefore,
$t_{cell} \ll t_{off}$ and we can rewrite Equation
\ref{eq:radiometercell} as

\begin{equation}
  t_{cell} = \frac{1}{\Delta\nu}\left(\frac{T_{sys}}{\sigma_{cell}}\right)^2 
  \label{eq:tcell}
\end{equation}

The integration time in a map cell is a function of the scanning rate
and the number of coverages of the map field, and should be calculated
with respect to the time accumulated in a Nyquist sampling cell.  For
illustration we assume a raster scanning geometry.  If we define
$\theta_x$ to be the angular distance along the scanning row,
$\theta_{ramp}$ to be a ``ramp-up'' distance at the beginning and end
of each row, and R to be the scanning rate (see \S\ref{scanrat}), then
the time to scan one row is

\begin{equation}
  t_{row} = \frac{\theta_x + 2\theta_{ramp}}{R}
  \label{eq:trow}
\end{equation}

\noindent{If} we also define $\theta_y$ to be the angular height of
the map and $\theta_{row}$ to be the angular sampling interval between
scanning rows, then the number of rows in the map is given by

\begin{equation}
  N_{row} = \frac{\theta_y}{\theta_{row}}
  \label{eq:nrow}
\end{equation}

\noindent{The} number of independent sampling cells in the map is

\begin{equation}
  N_{cell} = \frac{\left(\theta_x +
  2\theta_{ramp}\right)\times\theta_y}{\theta_N^2}
  \label{eq:ncell}
\end{equation}

\noindent{where} we are approximating the cells as rectangular.  The
integration time per cell, $t_{cell}$, is related to the above three
quantities as follows

\begin{equation}
  t_{cell} = \frac{t_{row} N_{row}}{N_{cell}}\times\eta
  \label{eq:tcell2}
\end{equation}

\noindent{where} $\eta$ is a convolution function correction factor
which takes account of the fact that we are using a tapered function
(a Gaussian-tapered Bessel function) instead of a pure sinc.  We
define the convolution function $C(u)_m$ as follows:

\begin{equation}
\label{eq:c}
C(u)_m = \int_0^{\infty} C(u) du
\end{equation} 

\noindent{The} $\eta$ correction factor is then given by

\begin{equation}
\eta \equiv \frac{C(u)_m}{C(0)_m}
\label{eq:convcor}
\end{equation}

\noindent{for} the truncated functions used in the AIPS task SDGRD
(see \S\ref{gridding} and Table \ref{tab:convfunc}).  Combining
Equations~\ref{eq:trow}, \ref{eq:nrow}, \ref{eq:ncell},
\ref{eq:tcell2}, and \ref{eq:convcor}, we find that 

\begin{equation}
t_{cell} = \frac{3.03 \theta_N^2}{R \theta_{row}}
\end{equation}

\noindent{Combining} this relation for $t_{cell}$ with our Radiometer
Equation (Equation~\ref{eq:radiometercell}), we find that the RMS noise
per cell in an OTF map is given by

\begin{equation}
\sigma_{cell} =
  \frac{T_{sys}}{\theta_N}\left(\frac{R\theta_{row}}{3.03 \Delta\nu}\right)^\frac{1}{2}
  \label{eq:rms}
\end{equation}

\noindent{Note} that a given RMS noise per cell obtainable with
multiple coverages and/or polarizations is given by

\begin{equation}
  \sigma = \frac{\sigma_{cell}}{\sqrt{N_c}}
  \label{eq:sigma}
\end{equation}

\noindent{where} $N_c$ is the number of coverages and/or polarizations
you combine to produce your final image.  Note that you can also
improve your map RMS by spatially smoothing it after gridding (see
\S\ref{gridding}). 

\subsubsection{Total Map Time}

Again assuming a raster scanning geometry, the total time required to
acquire a map must include not only the integration time scanning
across the field but also the time for calibration and OFF
integrations.  This can be written as

\begin{equation}
  t_{tot} = N_{row}\left[t_{row} + \frac{t_{off}}{N_{rpo}} +
\frac{t_{cal}}{N_{rpo}N_{opc}} + \frac{\epsilon}{N_{rpo}}\right]
\label{eq:maptime}
\end{equation}

\noindent{where} $N_{row}$ is the number of rows in the map, $t_{row}$
is given by Equation~\ref{eq:trow} above, $t_{off}$ is the OFF
integration time, $t_{cal}$ is the calibration integration time (\ie\
amplitude calibration and sky sample times), $N_{rpo}$ is the number
of rows per OFF measurement you do, $N_{opc}$ is the number of OFF
measurements per calibration measurement (amplitude calibration and
sky) you make, and $\epsilon$ is the ``overhead'' time, which is the
time that the telescope spends doing things other than integrating
(like moving from OFF positions to the map field).  The value for
$\epsilon$ often depends mainly on how far away the OFF position is
from the map, but 10 seconds is probably a good average 
estimate for vintage-1980 millimeter radio telescopes.  Most of the
time the amplitude calibration measurement is taken at the same
position and just before the OFF measurement, and so doesn't involve
additional overhead.  There may be other small overhead losses in
moving from row to row and starting scans, but these losses can
usually be neglected.

\subsection{Integration Time and RMS Noise}

Based on the following definitions:

\begin{description}
\item{$\eta\equiv$} Convolution function factor.  This is a measure of
how much the data oversamples the convolution function.
\item{$t_{row}\equiv$} Time to scan one row in the image.
\item{$N_{row}\equiv$} Number of rows in image.
\item{$\theta_{row}\equiv$} Row separation.
\item{$R\equiv$} Scanning rate.
\item{$N_{cell}\equiv$} Number of independent sampling intervals in
image.
\item{$\theta_{Nyquist}\equiv$} Nyquist sampling interval
($=\lambda$/(2D)).
\item{$T_{sys}\equiv$} System temperature.
\item{$\Delta\nu\equiv$} Spectral resolution.
\end{description}

\noindent{we} can write the relation for the integration time per
sampling interval as:

\begin{eqnarray}
t_{cell} &=& \frac{\eta t_{row} N_{row}}{N_{cell}} \nonumber \\
         &=& \frac{\eta \theta^2_{Nyquist}}{R \theta_{row}}
\end{eqnarray}

\noindent{and} the RMS noise per sampling interval is:

\begin{equation}
\sigma_{cell} = \frac{T_{sys}}{\theta_N}\left(\frac{R
  \theta_{row}}{\eta \Delta\nu}\right)^\frac{1}{2}
\end{equation}

The factor $\eta$ is dependent upon the convolution function used to
grid the OTF data.  The $\eta$ factors for the convolution
functions available in the AIPS single dish analysis programs are
listed in Table \ref{tab:convfunc}.

\begin{table*}
\centering
\caption{Convolution Function Factor}
\medskip
\label{tab:convfunc}
\begin{tabular}{|llccc|}
\hline
&& \multicolumn{2}{c}{\textsf{Function Form}} & \\
\textsf{Function} & \textsf{Parms} & \textsf{$\eta$ Linear} &
\textsf{$\eta$ Circular} & \textsf{$R_{support}$} \\
\hline
Pill Box &
\ldots & 1.00 & 0.78 & 0.5 \\
$\exp\left[-\left(\frac{|z|}{b}\right)^2\right]$ &
b = 1.00 & 3.14 & 3.14 & 3.0 \\
$\frac{\sin\left(\frac{\pi |z|}{a}\right)}{\frac{\pi |z|}{a}}$ &
a = 1.14 & 1.36 & 1.16 & 3.0 \\
$\frac{\sin\left(\frac{\pi |z|}{a}\right)}{\frac{\pi |z|}{a}}\times
{\exp\left[-\left(\frac{|z|}{b}\right)^2\right]}$ &
a = 1.55, b=2.52 & 2.33 & 1.43 & 3.0 \\
$\frac{J_1\left(\frac{|z|}{c}\right)}{\frac{|z|}{c}}\times
{\exp\left[-\left(\frac{|z|}{b}\right)^2\right]}$ &
b = 2.52, c = 1.55 & 3.59 & 3.03 & 3.0 \\
\hline
\end{tabular}
\end{table*}

\subsection{Differential Doppler Correction}

Since the correction for the radial motions of the local standard of
rest relative to a given source vary as a function of position on the
sky, time, and telescope location, the conversion from spectral
frequency to velocity varies over a given image.  This is a subtle
effect, traditionally ignored in spectral-line imaging.  However, for a
coordinate $1.4^\circ$ from the reference (\eg\ a $2^\circ$ square
field with a center reference), the error can be as large as 1.16 km
s$^{-1}$ (LSR velocities) or 0.79 km s$^{-1}$ (heliocentric) when
observed with the NRAO 12 Meter Telescope.  The LSR velocity error
changes significantly with time of year.  Narrow-band observations of
wide fields, such as observations of cold molecular regions, may be
seriously affected by this effect.  Fortunately, so long as the
spectra are fully sampled in frequency, this effect can be fully
corrected by doing the LSR calculations separately for every point
of each map, for all dates, and then interpolating to a common
frame and date in frequency space.

\subsection{Smoothing and Signal-To-Noise for Critically-Sampled Data}
\label{smoothing}

When you grid a critically-sampled (Nyquist-sampled) OTF map with a
sensible convolution function (such as 
the gaussian-tapered Bessel function described in \S\ref{gridding}),
the resulting spatial resolution is only minimally degraded.  However,
the \textit{noise} spectrum of your gridded map is flat (or,
equivalently, the noise in each gridded point is independent), while
the \textit{astronomy} spectrum follows that of the spatial frequency
response of the antenna.  If you now smooth the data significantly -- say to
a resultant beamwidth several times greater than the original spatial
resolution -- both the astronomical signal and the noise get smoothed.
The effective increase in integration time is given roughly by
$\frac{\textrm{new convolution function area}}{\textrm{original
    sampling cell area}}$. 
Obviously the signal-to-noise ratio increases by the square root of
this quantity.  Note that it is the ratio of the new
\textit{convolution function} to the
original \textit{sampling} that counts.  This is because the high
frequency terms of the original noise, which probably dominate the
total noise, are reduced in amplitude much more than the astronomical
signal, whose high-frequency components were already weighted down.

After you have smoothed the data once, the noise and astronomical signal now
have a very similar spectrum.  If you smooth the data a second time, again 
to a beamwidth several times larger than the current (smoothed) beamwidth, 
the effective integration time now increases roughly by
$\frac{\textrm{new beam area}}{\textrm{current beam
    area}}$.  It is now 
the ratio of the \textit{beams}, independent of the sampling of
either measurement.  This in a sense is because the noise 
in adjacent samples of the smoothed data is no longer independent, so you 
don't gain as much in the second smoothing as was gained in the first.

The conclusion is that to be rigorous you need to know the noise
spectrum as well as the data beamwidth in order to calculate what the
effective integration time becomes when you smooth critically-sampled
data.  It depends on whether you can consider noise in adjacent
samples independent.  After the initial gridding of a
critically-sampled data set, the noise \textit{is} independent in
adjacent samples.  After you have smoothed the data once, the noise is
no longer independent.

\section{OTF Observing at the NRAO 12 Meter Telescope}

Before starting a spectral line or continuum OTF map at the 12 Meter,
the telescope control system was configured to raster map the
target field.  For spectral line OTF, the map was taken in a total
power observing mode in the sense that you acquired a calibration spectrum
(a vane calibration) and a total power off measurement, followed by
one or more total power scanning rows, typically made up of hundreds
of individual spectra (the on measurements).  For continuum OTF, the
map was acquired using the continuum beam-switched observing mode.  In
this mode, the subreflector is switched between two positions (the
``+'' and ``--'' beam) in azimuth while the telescope scans.  Each of
the individual spectra or continuum ``+'' and ``--'' beam total power
measurements is tagged with the {\it actual} antenna encoder
positions.  As a result, antenna tracking errors caused by wind gusts,
for example, are actually recorded and taken into account in the data
analysis stage.

For spectral line OTF, the same calibration and off measurements were
used to calibrate all of the on measurements in the scanning rows
until another off measurement was taken.  Each total power off
measurement was given its own scan number.  All of the spectra or
continuum ``+'' and ``--'' beam total power measurements in each 
scanning row were concatenated along with arrays of time and position
information and stored on disk under a single scan number with a
single header.  The header information for each scanning row contained
the scan number of the previous off (for spectral line OTF) and
calibration measurements. 

Given the scanning parameters, the positioning system was configured
for tracking rates, row offsets, and the duration of a scanning row.
The spectral line data taking backend instructed the tracking system to
move to position and begin scanning.  Using some handshaking bits on
the telescope's status and monitor (SAM) bus, the two systems were
synchronized at the start of each row.  In addition, both the tracking
and backend computers had IRIG clock cards that were driven by the
observatory GPS standard clock.  The data backend read out the
spectrometer system every 100 ms for spectral line OTF.
The continuum data system was read out every 250 ms.  For
both spectral line and continuum OTF, the data backend tagged each data
parcel with the UT time stamp.  For the digital autocorrelation
spectrometer at the 12 Meter Telescope (called the Millimeter
Autocorrelator (MAC)), before the data could be read out Fourier
transforms of the 100 ms data samples needed to be
calculated.  To make data analysis as fast as possible, the FFT's were 
performed in real-time by the MAC control computers.  Every 10
ms, the tracking computer recorded its Az/El position with
respect to the field center.  Finally, at the end of the scanning row,
the position information from the tracking system was merged with the
data.  An interpolation of the position information was then made to
align slight differences between the time stamps of the two data sets.

\section{AIPS Processing}
\label{aips}

\begin{figure}
\resizebox{\hsize}{!}{
  \includegraphics[angle=-90]{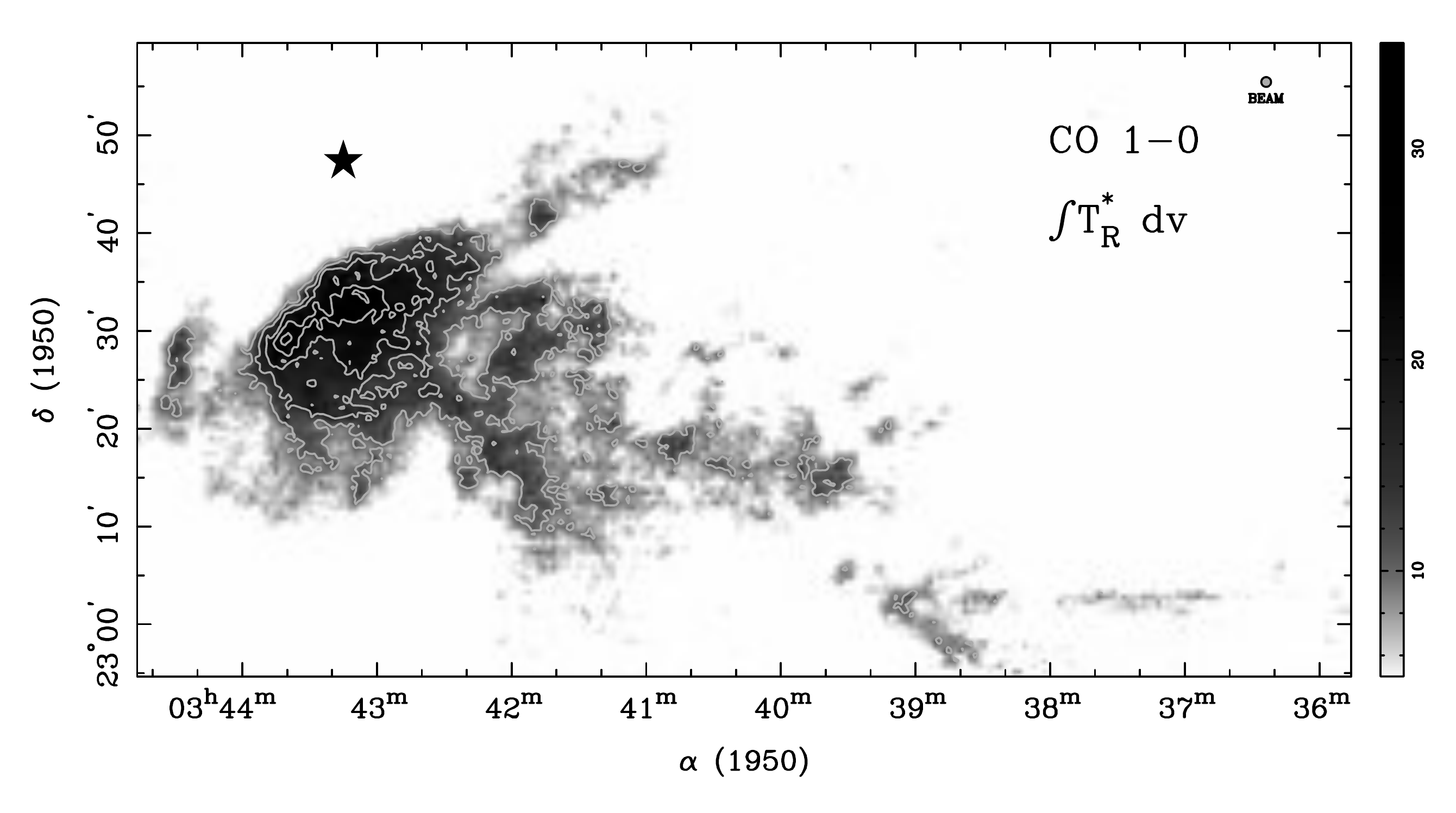}}
  \caption{CO $1\rightarrow0$ integrated intensity image of a
  molecular cloud near the Pleiades cluster.  The Pleiades member Merope
  is located in the upper left portion of this image, just off of the
  main emission component.  The CO emission appears to be tracing an
  interaction between the young stars in the Pleiades cluster and a
  nearby molecular cloud.}
  \label{fig:otfgallery}
\end{figure}

AIPS (\citet{Greisen2003}) was the default analysis software used to
process OTF data 
acquired with the NRAO 12m Telescope (see \citet{Mangum1999}).  The
main processing task was \textsf{SDGRD} (developed by Eric Greisen),
which read the raw 12m (``sdd'') data and gridded it using a variety
of user-selectable convolution functions.  Figure~\ref{fig:otfgallery}
shows a sample of an OTF image obtained with the NRAO 12 Meter
Telescope and processed with the AIPS OTF image analysis tasks.

\section{Conclusion}
\label{Conclusion}

The On-The-Fly (OTF) imaging technique is an efficient single-dish
radio telescope imaging mode that has been implemented at several
radio telescope observatories.  The relative gain in observing
efficiency is most extreme for the ``classic'' radio telescopes still
in operation.  This was certainly true for the NRAO 12 Meter
Telescope.  The NRAO 12 Meter Telescope began life as the 36 Foot
Telescope, the telescope responsible for the birth of
millimeter-wavelength molecular 
astronomy. Its history is one of success, innovation, and an
unparalleled desire by the staff operating this facility over its
32-year lifetime to provide a high level of service to the astronomical
community.

The 36 Foot Telescope made its first millimeter-wavelength astronomical
measurements in October 1968. This was the start of a period of
explosive growth in this new area of astronomical research, during which
most of the dozens of molecular species known to exist in the
interstellar medium were first detected with the 36 Foot.

In the early 1980s the telescope's reflecting surface and surface
support structure were replaced and in 1984 it was re-christened as the
12 Meter Telescope. Its scientific program subsequently evolved from one
dominated by astrochemistry to a broader mix of studies of molecular
clouds and Galactic star formation, evolved stars, astrochemistry, and
external galaxies.

The NRAO 12 Meter Telescope was the only millimeter-wavelength telescope in
the U.S. operated full-time as a national facility. More than 150
visitors used it each year. It offered users flexibility and the
opportunity to respond quickly to new scientific developments. Its
low-noise receiving systems covered a wide range of frequencies -- all
atmospheric windows from 68 GHz to 300 GHz -- and much attention was
given to making the instrument work reliably throughout this range.
Flexible spectral line and continuum back-ends allowed the observer to
match the instrument to the scientific goals. The development of
multi-beam receivers and the on-the-fly observing technique inaugurated
a new era of high-speed source mapping on angular scales complementary
to those of the millimeter-wave interferometers. The telescope control
system offered great flexibility and provided a proven remote observing
capability.

On February 22, 2000, NRAO announced that it would close the 12 Meter
Telescope at the end of the current observing season. On 26 July 2000,
the NRAO 12 Meter Telescope made its final astronomical measurements
as a U.S. national astronomical facility.  The telescope however
continues to be operated as part of the University of Arizona's
Steward Observatory.  The final NRAO data set was an 
on-the-fly image of the CO $1\rightarrow0$ emission from a star
formation region in the Cygnus-X region.

\end{document}